\documentclass{ws-procs9x6}

\begin{document}
\input epsf
\title{Branon dark matter: an introduction\footnote{
\uppercase{T}his work is supported by \uppercase{DGICYT} 
(\uppercase{S}pain)
under project numbers \uppercase{FPA} 2000-0956 and \uppercase{BFM} 
2002-01003}}

\author{J.A.R. Cembranos}

\address{Departamento de Estad\'{\i}stica e
Investigaci\'on Operativa III, and\\
 Departamento de  F\'{\i}sica Te\'orica,\\
 Universidad Complutense de
  Madrid, 28040 Madrid, Spain}

\author{A. Dobado and A.L. Maroto}

\address{Departamento de  F\'{\i}sica Te\'orica,\\
 Universidad Complutense de
  Madrid, 28040 Madrid, Spain}  

\maketitle

\abstracts{
This is a brief introduction to branon physics and its role
in the dark matter problem. 
We pay special attention to the phenomenological consequences,
 both in high-energy particle physics experiments and
in astrophysical and cosmological observations. }

\section{Introduction}
Most of the works done in the context of the brane-world 
scenario \cite{ADD} consider our world brane as a rigid
object which is placed at a given position in the extra
dimensions. However, rigid objects are incompatible with
Relativity and we should consider instead  branes as
dynamical objects which can move and fluctuate along the
extra dimensions. In such a case, apart from the Kaluza-Klein
(KK) modes of the gravitons propagating in the bulk space, new
fields appear on the brane which parametrize its position 
in the extra dimensions. This fields are the branons which
we will study in this work.

\section{Branon dynamics}
Let us consider  our four-dimensional space-time $M_4$ to be
embedded in a $D$-dimensional bulk space whose coordinates will be
denoted by $(x^{\mu},y^m)$, where $x^\mu$, with $\mu=0,1,2,3$,
correspond to the ordinary four dimensional space-time and $y^m$,
with $m=4,5,\dots,D-1$, are coordinates of the compact extra space
of typical size $R_B$. For simplicity we will assume that the bulk
metric tensor takes the following form:
\begin{eqnarray}
ds^2=\tilde g_{\mu\nu}(x)W(y)dx^\mu dx^\nu- g'_{mn}(y)dy^m dy^n
\label{metric}
\end{eqnarray}
where the warp factor is normalized as $W(0)=1$. The position of
the brane in the bulk can be parametrized as $Y^M=(x^\mu,
Y^m(x))$, and  we assume for simplicity that the ground state of
the brane corresponds to $Y^m(x)=0$.

In the simplest case in which the metric is not warped along the
extra dimensions, i.e. $W(y)=1$, the transverse brane fluctuations
are massless and they can be parametrized by the Goldstone boson
fields $\pi^\alpha(x),\; \alpha=4,5, \dots D-1$, associated to the
spontaneous breaking of the extra-space traslational symmetry.
 In that case we can choose the $y$
coordinates   so that the branon fields are proportional to the
extra-space coordinates: $\pi^\alpha(x) =f^2\delta_m^\alpha
Y^m(x)$, where the proportionality constant is related to the
brane tension $\tau=f^4$.

In the general case, the curvature generated by the warp factor
explicitly breaks the traslational invariance in the extra space.
Therefore branons  acquire a mass matrix which is given precisely
by the bulk Riemann tensor evaluated at the brane position:
$M^2_{\alpha\beta}=\tilde
g^{\mu\nu}R_{\mu\alpha\nu\beta}\vert_{y=0}$.

The dynamics of branons can be obtained from the
 Nambu-Goto action.
In addition, it is also possible to get their couplings to the
ordinary particles just by replacing the space-time by the induced
metric in the Standard Model (SM) action. Thus we get up to
quadratic terms in the branon fields \cite{DoMa,GB,BSky}:
\begin{eqnarray}
S_{Br} &=&\int_{M_4}d^4x\sqrt{\tilde g}\left[\frac{1}{2}
\left(\tilde g^{\mu\nu}\partial_{\mu}\pi^\alpha
\partial_{\nu}\pi^\alpha
-M^2_{\alpha\beta}\pi^\alpha \pi^\beta \right)\right.\nonumber \\
&+&\left.\frac{1}{8f^4}\left(4\partial_{\mu}\pi^\alpha
\partial_{\nu}\pi^\alpha-M^2_{\alpha\beta}\pi^\alpha \pi^\beta
\tilde g_{\mu\nu}\right)
T^{\mu\nu}_{SM}\right]
\label{Nambu}
\end{eqnarray}

We can see that branons  interact with the  SM
 particles through their energy-momentum tensor.
The couplings are controlled by the brane tension scale $f$. For
large $f$, branons are therefore weakly interacting particles. In
the case of a three-brane,
branons are pseudoscalar
particles. Parity on the brane then requires that branons always
couple to SM particles by pairs, which ensures that they are
stable particles. This fact means that branons are natural dark
matter candidates \cite{CDM}. 

\section{Limits from colliders}
\begin{figure}[h]
\centerline{\epsfxsize=5cm\epsfbox{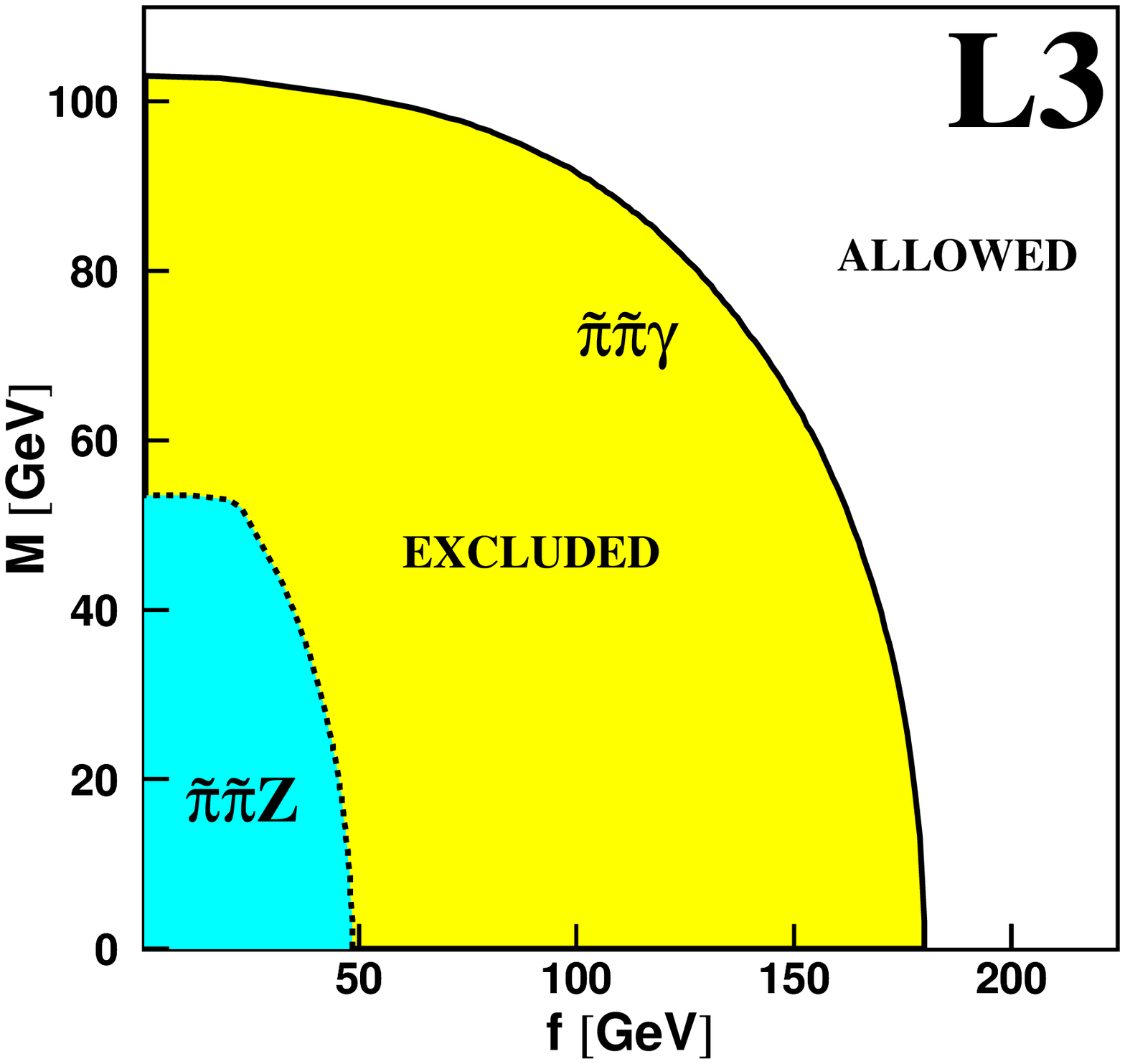}
\epsfxsize=6.5cm\epsfbox{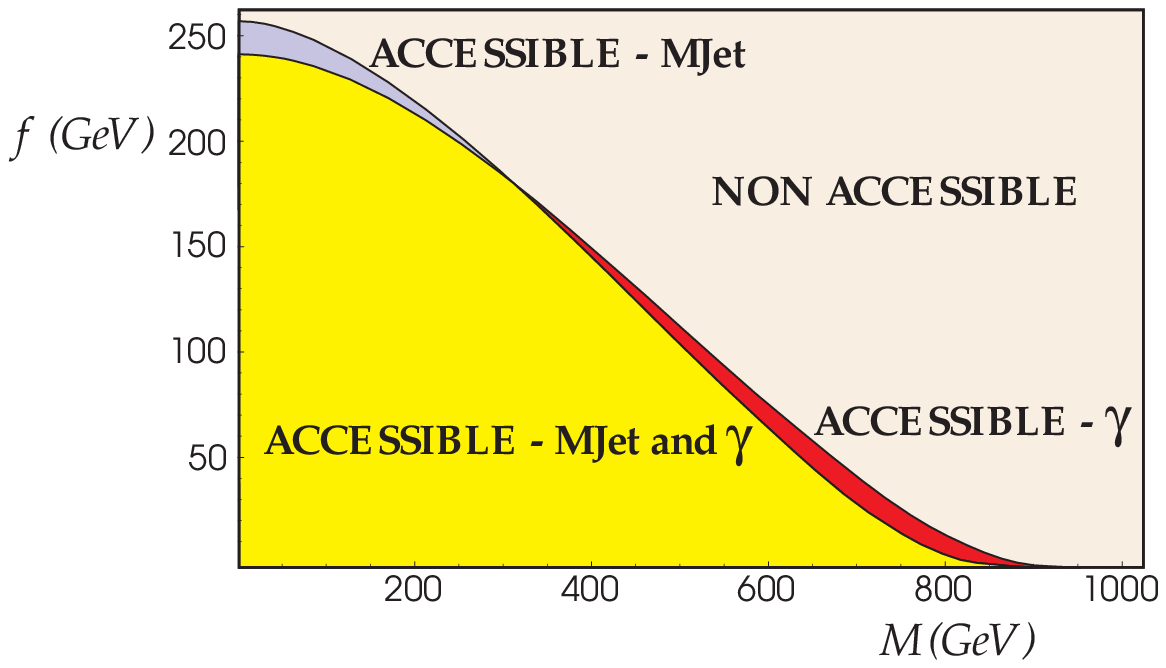}
}
\vspace{.3cm}
\footnotesize {\bf Figure 1:} 
Collider limits on branon parameters from single-photon and
single-Z processes at LEP (L3)  \cite{L3} (left). Limits from monojet 
and single-photon processes at Tevatron-I \cite{ACDM} (right)
\end{figure}
Collider experiments can be used to set bounds on the parameters
of branon physics, i.e. the brane tension scale $f$ and
the branon mass $M$. The L3 collaboration at LEP experiment
has recently obtained very stringent limits from the
analysis of single-photon processes in $e^+e^-$ collisions 
(see Fig. 1). In addition, we have also estimated the limits
coming from mono-jet and single-photon processes at Tevatron
(see Fig. 1).

\section{Cosmological and astrophysical limits}

\begin{figure}[h]
\centerline{\epsfxsize=10cm\epsfbox{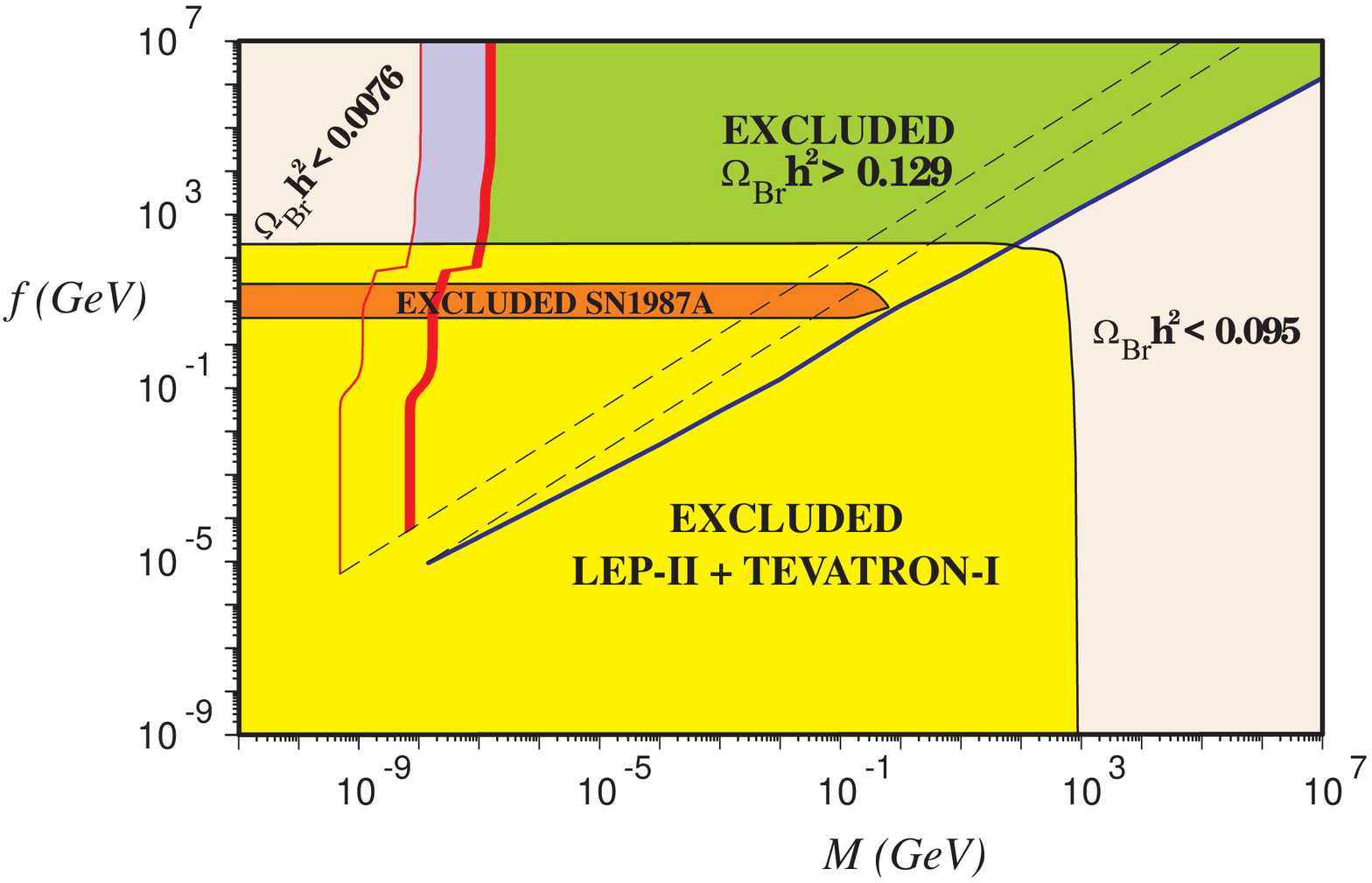}} 
{\footnotesize {\bf Figure 2:} 
Relic abundance in the $f-M$ plane for a model with one branon of
mass $M$. The two lines on the left correspond to the
$\Omega_{Br}h^2=0.0076$ and $\Omega_{Br}h^2=0.129 - 0.095$ curves
for hot-warm relics, whereas the right line corresponds to the
latter limits for cold relics (see \cite{CDM} for details). The
lower area is excluded by single-photon processes at LEP-II
\cite{ACDM} together with monojet signal at Tevatron-I
\cite{ACDM}. The astrophysical constraints are less restrictive
and they mainly come from supernova cooling by branon emission
\cite{CDM}.} 
\end{figure}
 
The potential WIMP nature of branons means that
these new particles are natural dark matter candidates.  In
\cite{CDM} the relic branon abundance has been calculated
in two cases: either
relativistic branons at freeze-out (hot-warm) or non-relativistic
(cold), and assuming that the evolution of the universe is
standard for $T<f$ (see Fig. 2). 
Furthermore, if the maximum
temperature reached in the universe is smaller than the branon
freeze-out temperature, but larger than the explicit symmetry
breaking scale, then branons can be considered as massless
particles decoupled from the rest of matter and radiation. In such
a case, branons can act as nonthermal relics establishing a
connection between the
coincidence problem and the existence of large extra dimensions
\cite{NT}.

\begin{figure}[ht]
\centerline{\epsfxsize=8cm\epsfbox{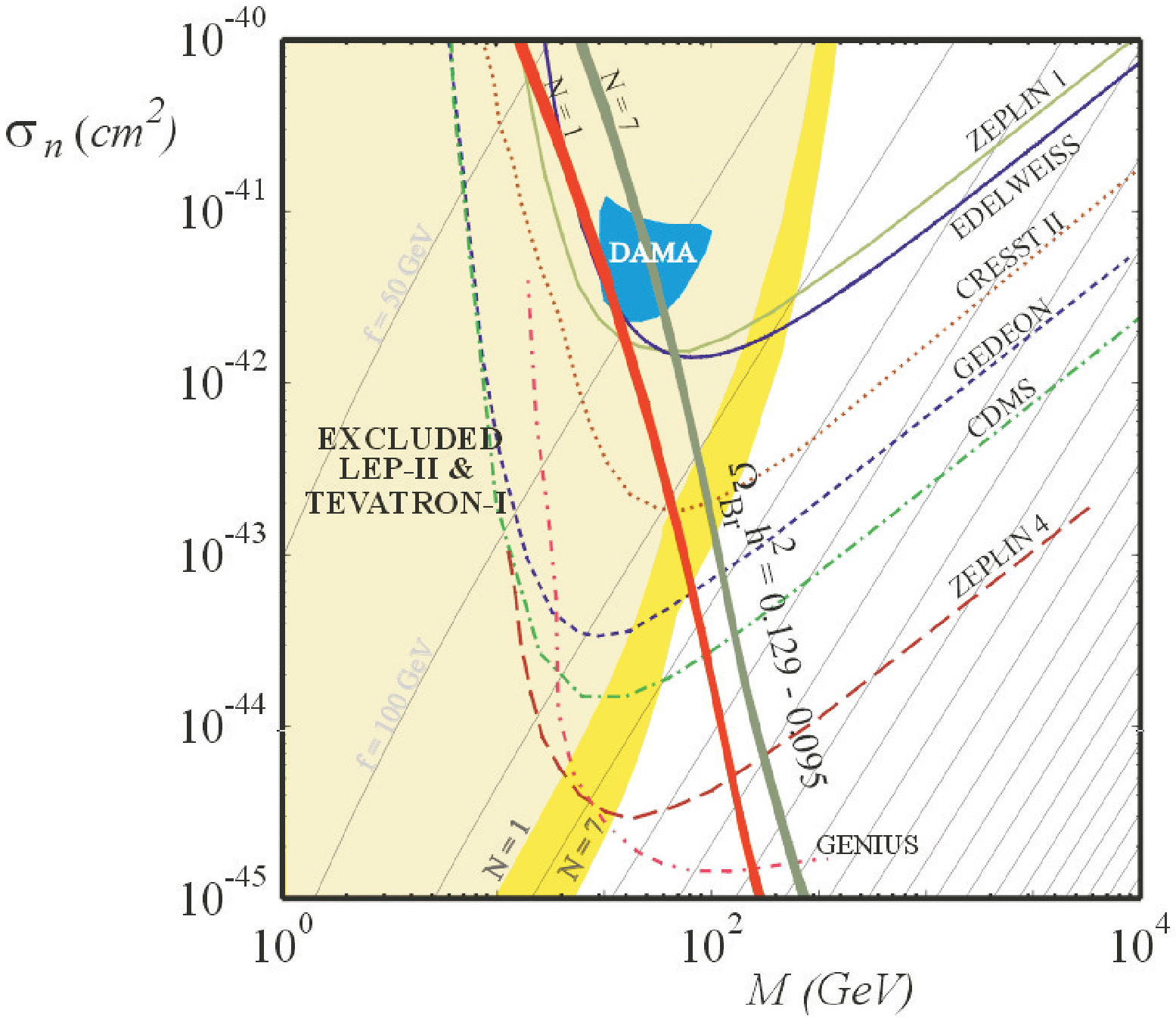}} 
{\footnotesize {\bf Figure 3:} Elastic branon-nucleon cross section $\sigma_n$ in terms
of the branon mass. The thick (red) line corresponds to the
$\Omega_{Br}h^2=0.129-0.095$ curve for cold branons in Fig. 2 
from $N=1$ to $N=7$. The shaded areas are the LEP-II
and Tevatron-I exclusion regions. The
solid lines correspond to the current limits on the
spin-independent cross section from direct detection experiments.
The discontinuous lines are the projected limits for future
experiments. Limits obtained from \cite{dmtools}.}
\end{figure}
If branons make up the galactic halo, they could be detected by
direct search experiments from the energy transfer in elastic
collisions with nuclei of a suitable target. From
Fig. 3 we see that if branons constitute the
dominant dark matter component, they could not be
detected by present experiments such as DAMA, ZEPLIN 1 or
EDELWEISS. However, they could be observed by future detectors
such as CRESST II, CDMS or GENIUS \cite{CDM}.

Branons could also be detected indirectly: their annihilations in
the galactic halo can give rise to pairs of photons or $e^+ e^-$
which could be detected by $\gamma$-ray telescopes such as MAGIC
or GLAST or antimatter detectors (see \cite{CDM} for an estimation
of positron and photon fluxes from branon annihilation in AMS).
Annihilation of branons trapped in the center of the sun or the
earth can give rise to high-energy neutrinos which could be
detectable by high-energy neutrino telescopes such as AMANDA,
IceCube or ANTARES. These searches complement those in high-energy
particle colliders (both in $e^+ e^-$ and hadron colliders
\cite{L3,ACDM}) in which real (see Fig. 1) and virtual branon
effects could be measured \cite{CDM}. Finally, quantum fluctuations
of branon fields during inflation can give rise to CMB anisotropies
through their direct contribution to the induced metric (work is
in progress in these directions).

\end{document}